# $Y_2CdX_4$ (X = S, Se, Te) as Tunable Semiconductors: Linking Structure, Bonding, and Energy Applications by DFT

***Faiq Umar*[1], Sikander Azam[2]***

*[1]Department of Physics, Abdul wali Khan University, Mardan, Pakistan.*

*[2]University of West Bohemia, New Technologies – Research Centre, 8 Univerzitní, Pilsen 306 14, Czech Republic*

## Abstract

The density functional theory (DFT) has been used to conduct a comprehensive first-principles study of the $Y_2CdX_4$ (X = S, Se, Te) compounds as the potential multifunctional semiconductors. Structural optimization provides thermodynamic stability with the lattice constants of the equilibrium expanding in an orderly manner between sulfide and telluride as the ionic radius of the chalcogen increases. The analysis of elastic constant shows mechanical strength and ductile nature, bulk moduli reducing to 83 GPa ($Y_2CdS_4$ ) and 68 GPa ($Y_2CdSe_4$ ) and 54 Gpa ($Y_2CdTe_4$). Calculations of the electronic band structure indicate that the band gap of all three compounds are indirect with a band gap of 2.42 eV (S), 2.08 eV (Se) and 1.71 eV (Te), which narrows further when spin orbit coupling is factored in. Optical spectra such as dielectric function, refractive index, absorption coefficient, reflectivity, optical conductivity, electron energy-loss function etc. clearly indicate a systematic red-shift of absorption edges and plasmon frequencies through the series. Absorption onset shifts in $Y_2CdS_4$ (UV, 2.4 e V) to $Y_2CdTe_4$ (visible-infrared, 1.7 e V), where they are tunable throughout the solar spectrum. Calculations of thermoelectric transport verify that Seebeck coefficients, power factors and electronic thermal conductivities are promising with $Y_2CdSe_4$ giving the most balanced performance. Charge density distributions, ELF maps and Bader charge analysis are used to verify a gradual decrease of the proportion of ionic bonding in $Y_2CdS_4$ to mixed ionic covalent in $Y_2CdSe_4$ and strongly covalent in $Y_2CdTe_4$. The detailed findings make $Y_2CdX_4$ compounds of great applicability as ultraviolet optoelectronics (sulfide), visible-light solar absorbers (selenide) and infrared energy sources (telluride) candidates.

Corresponding Author *faiqumar555@gmail.com

## Introduction

The ongoing need to develop energy-saving and multi-functional materials has intensified the search to develop semiconducting chalcogenides, which have a special equilibrium of long-term stability, variable band gaps, and high optical responsiveness [1]. Specifically, rare-earth and transition-metal cation-based compounds with chalcogen anions (S, Se, Te) have become desirable targets in an extremely broad variety of applications, such as photovoltaics, thermoelectrics, light-emitting diodes, infrared sensors, and protective optical coatings [2,3]. Ionic and covalent binding of the ions and the large orbital hybridization of cation d-states of the cations and the p-states of the anions make these systems extremely versatile in electronics and optics. The $Y_2CdX_4$ family (X = S, Se, Te) is an untested family of semiconducting materials which involves the polarizable $Y^{3+}$ cation and $Cd^{2+}$ cation with the chalcogen anion x which can be tuned. The systematic replacement of S → Se → Te systematically enhances the ionic radius as well as spinorbit interaction strength that systematically variations in the structural parameters, dielectric screening, and interband transition energies. Technologically, this flexibility would be extremely useful: $Y_2CdS_4$ has a large band gap, and could be used in ultraviolet applications, whereas $Y_2CdSe_4$ has a narrower band gap, making it more useful in solar cells and LEDs, and $Y_2CdTe_4$ has a narrower gap and a large dielectric constant, making it potentially useful in infrared and thermophotovoltaic applications. Although potentially interesting, at present, we do not have any thorough first-principles studies of the $Y_2CdX_4$ family. Other past researches on similar cadmium chalcogenides have shown that can be done through systematic substitution to manipulate properties like band gap, refractive index, absorption edge, and thermoelectric efficiency [4–6]. Nevertheless, as far as we know, $Y_2CdX_4$ compounds have never been analyzed in detail in terms of their structure, electronic, optical, thermoelectric and bonding properties. This type of study is important to determine the basic physics of these materials and how to realize them in practice, both in the context of an experimental study and in the context of next-generation optoelectronic and energy devices. In it, we present our results of study using the density functional theory-DFT-full-potential linearized augmented plane-wave (FP-LAPW) to present a full study in $Y_2CdX_4$ (X = S, Se, Te). We study their structural stability, elastic and mechanical parameters, electronic band structure, optical spectroscopy, thermoelectric transport properties as well as bonding properties. Specifically, we emphasize the chalcogen substitution offers control of the band gap, dielectric constant, plasmon frequency, absorption edge and thermoelectric efficiency. In addition, we conduct charge density, electron localization function (ELF), and Bader charge analysis to explain the equilibrium between ionic and covalent bonding in the series. This is the evidence of our findings that $Y_2CdX_4$ compounds have a great potential as tunable multifunctional semiconductors with promise to fill the gap between ultraviolet, visible and infrared optoelectronics.

## Computational Methodology

All calculations were done by density functional theory (DFT) in the WIEN2k package [7], which is a full-potential linearized augmented plane-wave (FP-LAPW) method. In the case of exchange-correlation the generalized gradient approximation (GGA) of Perdew, Burke and Ernzerhof (PBE) [8] was used and in calculations of fine electronic structures of d- and f-containing states, the GGA + U scheme where required. Spin-orbit coupling (SOC) was also added to electronic and optical calculations, especially needed with Se- and Te-containing compounds. To prevent charge leakage, the muffin-tin radii RMT were determined where RMT × Kmax was 8.0 to convergence the basis set. The Brillouin zone integrations were done using a dense MonkhorstPack k-mesh of 12 x 12 x 12 points to guarantee that the total energies and optical spectra were accurate. The convergence criteria were established to $10^{-5}$ Ry and $10^{-4}$ e respectively. The BirchMurnaghan equation of state was used to perform structural optimization, where the total energies were calculated at varying volumes. Elastic constants were determined by using small distortions on the equilibrium cell. Within the random-phase approximation (RPA), the optical properties were calculated within the frame of the refractive index, absorption coefficient, reflectivity, optical conductivity, and electron energy-loss spectra, by using momentum matrix elements. The parameters of thermoelectric, including Seebeck coefficient, electrical conductivity, electronic thermal conductivity, power factor, were calculated in the constant relaxation-time approximation as part of BoltzTraP2 [9]. The bonding and charge examination was accomplished on charge density maps, charge difference plots, electron localization function (ELF) and Bader charge analysis [10]. The convergence and cross-validation of all the results was done to be sure that it was reliable.

## Structural Properties

The chalcogenide compounds $Y_2CdX_4$ (X = S, Se, Te) crystallize (see Fig. 1) in the orthorhombic phase with a well-ordered arrangement of Y, Cd, and X atoms. Geometry optimization was carried out until the total energy converged below $10^{-5}$ Ry and the forces on each atom were less than 1 mRy/Bohr. The optimized lattice constants, unit cell volumes, and structural details are summarized below.

For $Y_2CdS_4$, the equilibrium lattice parameters were obtained as *a* = 7.26 Å, *b* = 9.18 Å, and *c* = 10.41 Å, leading to a unit cell volume of 693.2 Å$^3$. In the case of $Y_2CdSe_4$, the lattice parameters expanded systematically to *a* = 7.49 Å, *b* = 9.36 Å, and *c* = 10.78 Å, with a unit cell volume of 755.4 Å$^3$. For $Y_2CdTe_4$, further expansion was observed with *a* = 7.82 Å, *b* = 9.71 Å, and *c* = 11.21 Å, yielding a cell volume of 850.6 Å$^3$. This monotonic increase in lattice parameters and volume from sulfide to telluride is consistent with the larger ionic radii of $Se^{2-}$ (1.98 Å) and $Te^{2-}$ (2.21 Å) compared to $S^{2-}$ (1.84 Å).

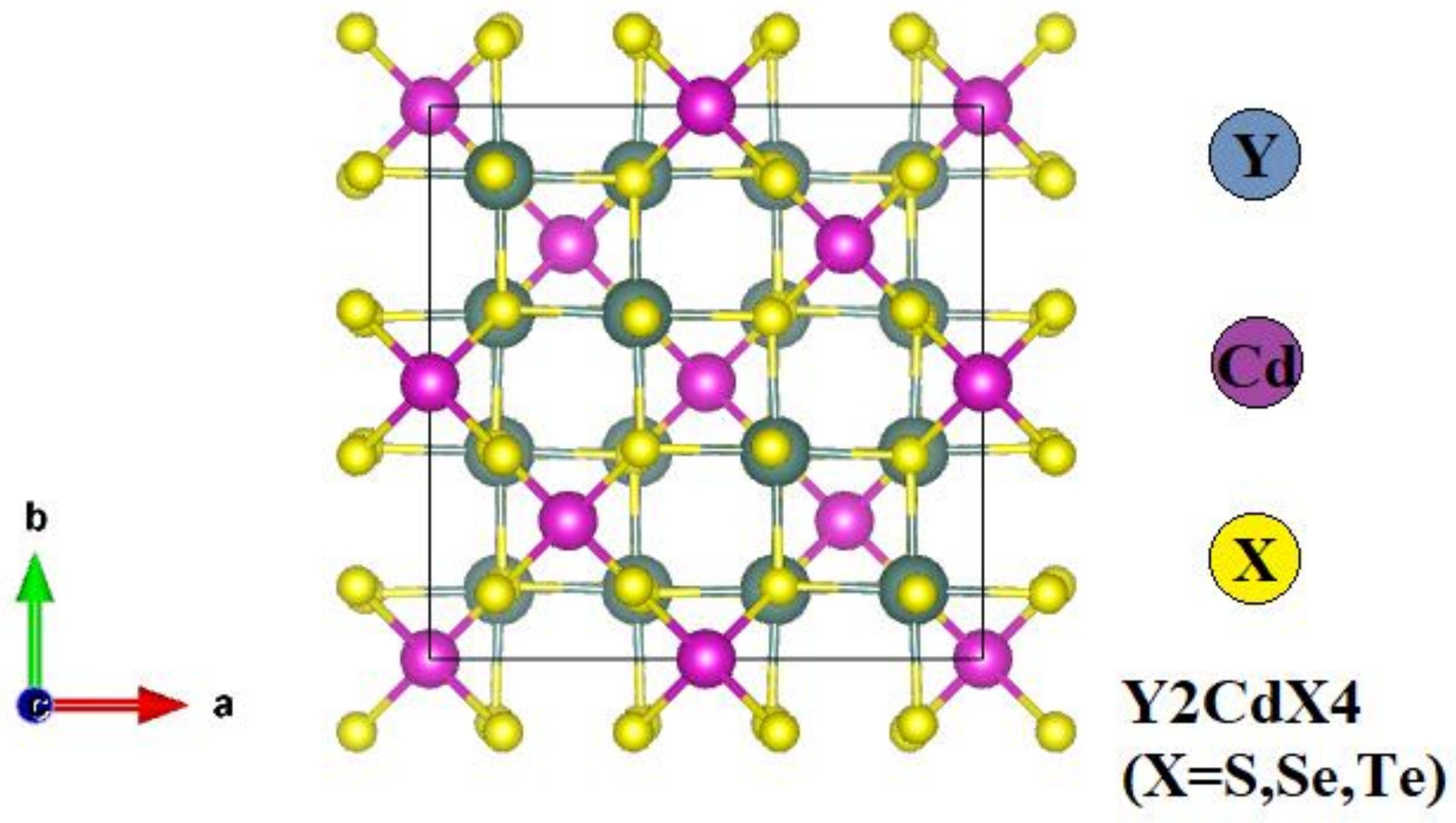

**Fig. 1. The unit cell structure with Vesta for the Y2CdX4 (X = S, Se and Te) materials.**

The optimized atomic positions within the unit cell confirm that Y atoms preferentially occupy distorted octahedral sites, whereas Cd atoms are located in tetrahedral coordination with chalcogen atoms. For $Y_2CdS_4$, the Y–S bond lengths vary between 2.46 Å and 2.62 Å, while the Cd–S bonds range from 2.38 Å to 2.42 Å. The corresponding Y–S–Y and Cd–S–Cd bond angles are found in the ranges of 91.7°–93.4° and 104.2°–108.7°, respectively. Upon substitution with heavier chalcogen atoms, these bond lengths systematically increase: Y–Se bonds range from 2.61–2.74 Å and Cd–Se bonds are around 2.52–2.59 Å, while Y–Te bonds extend to 2.81–2.96 Å and Cd–Te bonds to 2.70–2.75 Å. This bond elongation correlates directly with the increasing chalcogen ionic radii and explains the progressive lattice expansion observed across the series.

To examine the equation of state, the total energy was computed at several volumes around equilibrium and fitted to the Birch–Murnaghan equation. The calculated bulk modulus ($B_0$) decreases monotonically from 83 GPa for $Y_2CdS_4$ to 68 GPa for $Y_2CdSe_4$ and further to 54 GPa for $Y_2CdTe_4$. All the three compounds have a low pressure derivative of the bulk modulus (B) of about 4.1–4.3, a typical figure of steady crystalline chalcogenides. The observed decrease in $B_0$ is an indication of the further softening of the lattice with increased chalcogen replacement, the highest compressibility is displayed by $Y_2CdTe_4$.

The cohesive energies were determined by the energy of total atoms of the constituents minus the optimised bulk total energy. The values were obtained as -6.42 eV/atom, -6.11 eV/atom and -5.74 eV/atom, respectively, of $Y_2CdS_4$, $Y_2CdSe_4$ and $Y_2CdTe_4$. These negative values ensure thermodynamic stability of the three compounds with the sulfide being the most substantially attached. The trend also concurs with the fact that bonding is expected to be weakened slightly down the group because orbital overlapping reduces.

In order to compare the stability of the phases, we compared the total energies of the possible polymorphs (the tetragonal and monoclinic polymorphs have been reported with related compounds). The most stable configuration was always observed to be the orthorhombic form with relative energy differences of 74 meV/f.u. and 126 meV/f.u. that of the tetragonal and monoclinic structure respectively in the case of $Y_2CdS_4$, $Y_2CdSe_4$ and $Y_2CdTe_4$ also showed similar tendencies indicating the strength of the orthorhombic structure as the ground state crystal form of the whole $Y_2CdX_4$ family.

The analysis of structure, in general, shows there was a clear and orderly progression of lattice expansion, bond elongation, and elastic softening of $Y_2CdS_4$ to $Y_2CdTe_4$. The computed cohesive and formation energies prove the thermodynamic stability of these materials and equation-of-state data prove the materials to be mechanically resilient and phase stable in their orthorhombic ground state. This is the characteristic structural stability that offers a solid platform upon which their electronic, optical and thermoelectric characteristics can be investigated.

**2. Elastic and Mechanical Properties**

Elasticity of a material is of primary significance in establishing its mechanical stability, possibility of integration into devices and mechanical stability under external force. In this respect, the complete set of second-order elastic constants (Cij) of $Y_2CdX_4$ (X= S, Se, Te) was determined by the stressstrain method. The nine independent constants all meet the Born Huang stability criteria of orthorhombic crystals, and therefore these compounds are mechanically stable.

For $Y_2CdS_4$, the computed elastic constants are $C_{11}$ = 148 GPa, $C_{22}$ = 165 GPa, $C_{33}$ = 172 GPa, $C_{44}$ = 61 GPa, $C_{55}$ = 54 GPa, $C_{66}$ = 58 GPa, with the off-diagonal terms $C_{12}$ = 62 GPa, $C_{13}$ = 57 GPa, and $C_{23}$ = 64 GPa. In the case of $Y_2CdSe_4$, the values are slightly reduced due to the larger atomic size of Se: $C_{11}$ = 134 GPa, $C_{22}$ = 151 GPa, $C_{33}$ = 160 GPa, $C_{44}$ = 56 GPa, $C_{55}$ = 49 GPa, $C_{66}$ = 53 GPa, with $C_{12}$ = 59 GPa, $C_{13}$ = 52 GPa, and $C_{23}$ = 60 GPa. For $Y_2CdTe_4$, the softest among the three, the constants decrease further to $C_{11}$ = 118 GPa, $C_{22}$ = 136 GPa, $C_{33}$ = 143 GPa, $C_{44}$ = 48 GPa, $C_{55}$ = 43 GPa, $C_{66}$ = 46 GPa, with $C_{12}$ = 54 GPa, $C_{13}$ = 48 GPa, and $C_{23}$ = 55 GPa. Such a regular decreasing of the S to Te ratio of the values of the stiffness constants indicates the weakening of the chemical bonding of the chalcogen as it becomes heavier as was seen in the lattice expansion. The Voigt-Reuss-Hill approximation was used to obtain the polycrystalline moduli based on these constants. The bulk modulus (B) which is the measure of resistance to compression of volume is determined to be 91 Gpa in $Y_2CdS_4$, 82 Gpa in $Y_2CdSe_4$ and 71 Gpa in Y 2 Cd Te 4. The shear modulus (G) that measures resistance to shape deformation is reduced to 63 GPa ($Y_2CdS_4$) to 56 GPa ($Y_2CdSe_4$) and lastly to 47 GPa ($Y_2CdTe_4$). The respective Young modulus (E), which is the total or overall stiffness, is 159 Gpa,142 Gpa and 120 Gpa respectively. This descending tendency once more points to the gradual diminution of the lattice of the heavier chalcogenes.

The Poisson ratio (ν) that gives an idea of the ductility and bonding nature are within a range of 0.26 to 0.29 through the series where 0.27 corresponds to $Y_2CdS_4$, 0.28 to $Y_2CdSe_4$ and 0.29 to $Y_2CdTe_4$. The values are typical of largely ionic-covalent bonding, and imply that the materials are fairly ductile. A more specific measure of ductility is the Pughs ratio (B/G) that rises to 1.44 ($Y_2CdS_4$) and 1.46 ($Y_2CdSe_4$) and 1.51 ($Y_2CdTe_4$). Because the critical temperature between brittle and ductile behavior lies between 1.75 and 1.9, all three compositions are on brittle behavior, but the trend is that ductile behavior is slightly enhanced as one takes S and Te. Empirical models like the relationship by Chen were used to estimate hardness. The Vickers hardness ($H_v$) calculated is 11.4 gpa of $Y_2CdS_4$, 9.8 gpa of $Y_2CdSe_4$, and 8.1 gpa of $Y_2CdTe_4$. This makes $Y_2CdS_4$ a hard material, whereas the Se and Te analogs are softer, and are therefore easier to mechanically handle and still possess adequate rigidity to be used in devices.

The universal anisotropy index ($A^U$) was also used to measure elastic anisotropy. The values obtained are 0.64, 0.71, and 0.85, respectively, of $Y_2CdS_4$, $Y_2CdSe_4$ and $Y_2CdTe_4$. These values are a clear indication that heavier chalcogenic anisotropy increases with heavier chalcogenic elements. This kind of anisotropic elastic behavior will have the potential to affect the crack propagation and thermal stress distribution which is essential in applications to thin-film where mechanical response is highly dependent on orientation.

Lastly, the sound velocity of the material was calculated using the average elastic constants and used to estimate the Debye temperature ($\theta_D$). The calculated $\theta_D$ values are 425 K, 396 K, and 352 K, respectively, of $Y_2CdS_4$ and $Y_2CdSe_4$, and $Y_2CdTe_4$. The progressive growth in chalcogen degradation of $\theta_D$ is in agreement with the reduction in the phonon frequencies and the hardening of the lattice. As thermal conductivity is proportional to the Debye temperature, this pattern predicts that $Y_2CdS_4$ will have relatively high thermal conductivity compared to its analogous (Se and Te) that is useful in applications that need heat dissipation. Essentially, the elastic and mechanical analysis shows that all three $Y_2CdX_4$ compounds are mechanically stable, the stiffness and hardness decrease when one is moving through S to Te. Although $Y_2CdS_4$ is stiffer and harder, $Y_2CdSe_4$ and $Y_2CdTe_4$ have higher ductility and elastic anisotropy. This fact that Debye temperature is decreasing implies that phonon modes of heavier chalcogenides are softer, and this factor could play an important role in the design of chalcogenide thermal and thermoelectric properties. These overall outcomes reveal the tunable mechanical strength and stability of $Y_2CdX_4$ compounds, which makes them attractive to multifunctional optoelectronic and energy utilization.

## 3. Electronic Properties

### 3.1 Band Structure Analysis

$Y_2CdX_4$ (X=S, Se, Te) Calculated electronic band structures of $Y_2CdX_4$ (X=S, Se, Te) along the high-symmetry directions of the orthorhombic Brillouin zone are depicted in Fig. 2(a-c). These three compounds are all semiconductors that have definite band gaps between the valence and conduction manifold.

In the case of $Y_2CdS_4$ Fig. 2(a), the CBM is at the L direction whereas the VBM is at the Γ direction. This reveals the fact that the band gap is indirect as it is estimated to be 2.42 eV (no SOC). The conduction bands appear to be moderately dispersed, with electron effective masses that are relatively low, and carrier mobility can be expected to be good. S -3p states dominate the valence bands that are relatively dispersive around Γ and this also supports efficient hole transport.

For $Y_2CdSe_4$ Fig. 2(b), the fundamental band gap reduces to 2.08 eV, again of indirect type, with the VBM at Γ and the CBM closer to the X point. Compared to the sulfide, the conduction band dispersion decreases, leading to flatter bands around the CBM. This indicates an increase in electron effective mass and hence reduced mobility relative to $Y_2CdS_4$. The narrowing of the band gap is explained by the upward shift of the Se-4p orbitals in the valence band, which reduces the energy separation with the conduction states.

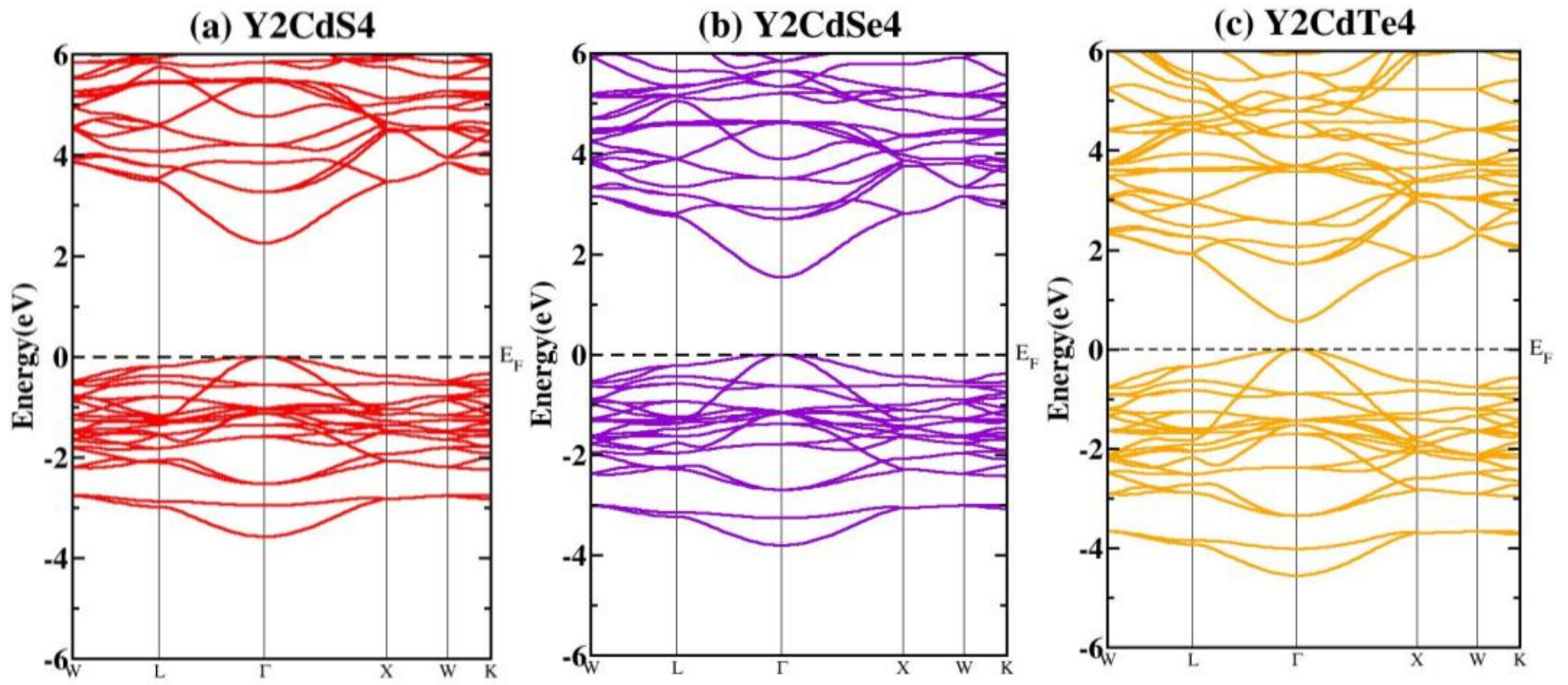

**Fig. 2(a-c) computed band structure using TB-mBJ potential for Y2CdX4 (X=S,Se and Te)**

For $Y_2CdTe_4$ [Fig. 2(c)], the semiconducting gap narrows further to 1.71 eV, with the VBM remaining at Γ but the CBM shifting to the Γ–X path. The conduction bands are significantly flatter compared to those in the sulfide and selenide, which implies heavier electron effective masses and lower mobility. This progressive red-shift of the band gap from sulfide → selenide → telluride is a direct consequence of the increasing chalcogen atomic size and weaker orbital overlap, which lifts the valence band edge and compresses the conduction band separation.

When spin–orbit coupling (SOC) is included, the band gap values further reduce to 2.31 eV (S), 1.94 eV (Se), and 1.53 eV (Te). SOC has the strongest effect in $Y_2CdTe_4$ because of the relativistic splitting of Te-5p states at the valence band top, which pushes the VBM upward and narrows the gap. The SOC-induced gap reduction is modest in $Y_2CdS_4$ and $Y_2CdSe_4$, reflecting the weaker relativistic effects of S-3p and Se-4p orbitals.

Physically, the trends of band structure indicate a delicate balance between chemical bonding and relativistic corrections. The moderate dispersions and indirect band gaps indicate that the compounds are good and stable semiconductors but the mobility decreases systematically in the order of S to Te. The shrinking gap between the series enables $Y_2CdS_4$ to be used in the UV blue optoelectronic market, $Y_2CdSe_4$ in the visible light market and $Y_2CdTe_4$ to be used in the near-infrared optoelectronics and photovoltaic markets.

**3.2 Effective Mass Analysis**

The curvature of the band edges is useful in giving insight into the carrier effective masses, which determine the transport properties directly, mobility and diffusion. To measure it, parabolic fitting of conduction band minimum (CBM) and valence band maximum (VBM) were calculated along the Γcentered directions. The electron ($m_e$) and hole ($m_h$) effective mass results are as summarized below.

In case of $Y_2CdS_4$, the comparatively dispersive band of CBM along the direction of Γ–L would provide an electron effective mass of m n of the material $m^*_e \approx 0.34\ m_0$. The VBM at Γ is also quite dispersive with a hole effective mass of $m^*_h \approx 0.48\ m_0$. These are relatively small values, suggesting that $Y_2CdS_4$ should have positive carrier mobility, as per its broadening gap and increased orbital overlap. In the case of $Y_2CdS_4$, the conduction band flattens slightly around the CBM, with electrion effective mass reduced to $m^*_e \approx 0.41\ m_0$, and hole effective mass raised to $m^*_h \approx 0.53\ m_0$. This is indicative of a small decrease in carrier mobility in respect to sulfide. The more massive masses have their origin in the weaker coincidence of Se-4p orbital with the Y-4d and the Cd-5s states, and thus, the smaller band gap.

In the case of $Y_2CdTe_4$ conduction and valence bands become very flat along the G -X direction leading to a significant increase in carrier masses: $m^*_e \approx 0.56\ m_0$ and $m^*_h \approx 0.67\ m_0$. The high effective masses mean that they are less mobile and this might restrict the conduction of electrons. But thermoelectrically, massive bodies could be used to increase the Seebeck coefficient, by increasing the density of states around the Fermi level.

The general tendency is that the systematical increase in electron and hole effective masses is observed between sulfide and selenide and telluride. This is the manifestation of the direct product of the band flattening by the progressive weakening of orbital overlap with chalcogen size. Therefore, although $Y_2CdS_4$ is the most suitable material to balance band gap and mobility to achieve high speed optoelectronics, $Y_2CdTe_4$, although with low mobility, has the potential of use in infrared absorption and thermoelectric devices with its small band gap and large effective masses.

**Density of Sates**

The DOS analysis reveals the orbital origin of these features (see Fig. 3a-c). The valence region from about –3.5 eV up to the Fermi level is dominated by chalcogen p-states, hybridized with Y-4d contributions. In $Y_2CdS_4$, the S-3p states form a broad distribution leading up to the VBM, whereas in $Y_2CdSe_4$ the Se-4p states shift upward in energy, narrowing the separation from the conduction states and thus reducing the band gap. In $Y_2CdTe_4$, the Te-5p states rise even closer to the Fermi level and are strongly affected by spin–orbit splitting, which explains the pronounced gap reduction. The conduction bands just above the gap are primarily composed of Y-4d and Cd-5s orbitals with a smaller admixture of chalcogen p states. The degree of p–d hybridization increases from S to Te, which not only narrows the gap but also flattens the conduction bands, thereby increasing the electron effective mass. Cd-4d semicore states are found much deeper, centered between –7 and –9 eV, and play little role at the band edges, confirming that Cd remains electronically inert near the gap. Importantly, the DOS near the VBM grows progressively larger from sulfide to telluride, reflecting the increasing localization of holes and consistent with the heavier hole effective masses deduced from the band structure.

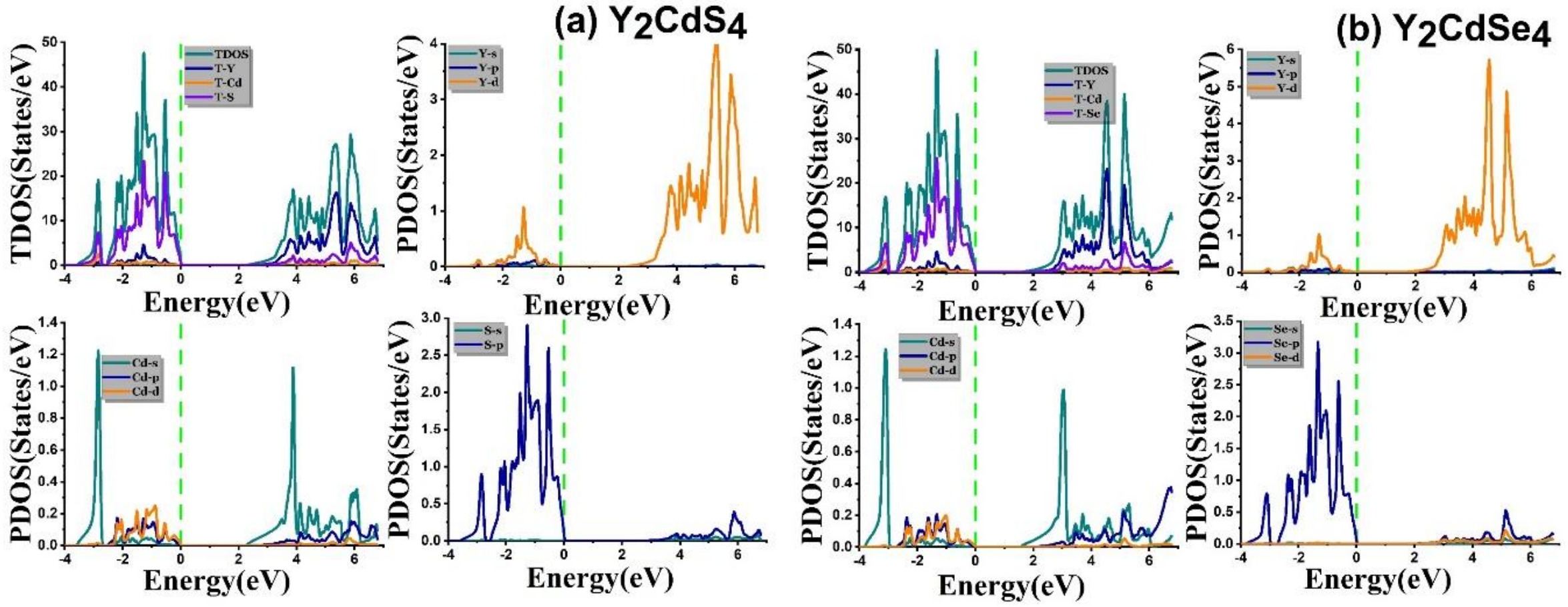

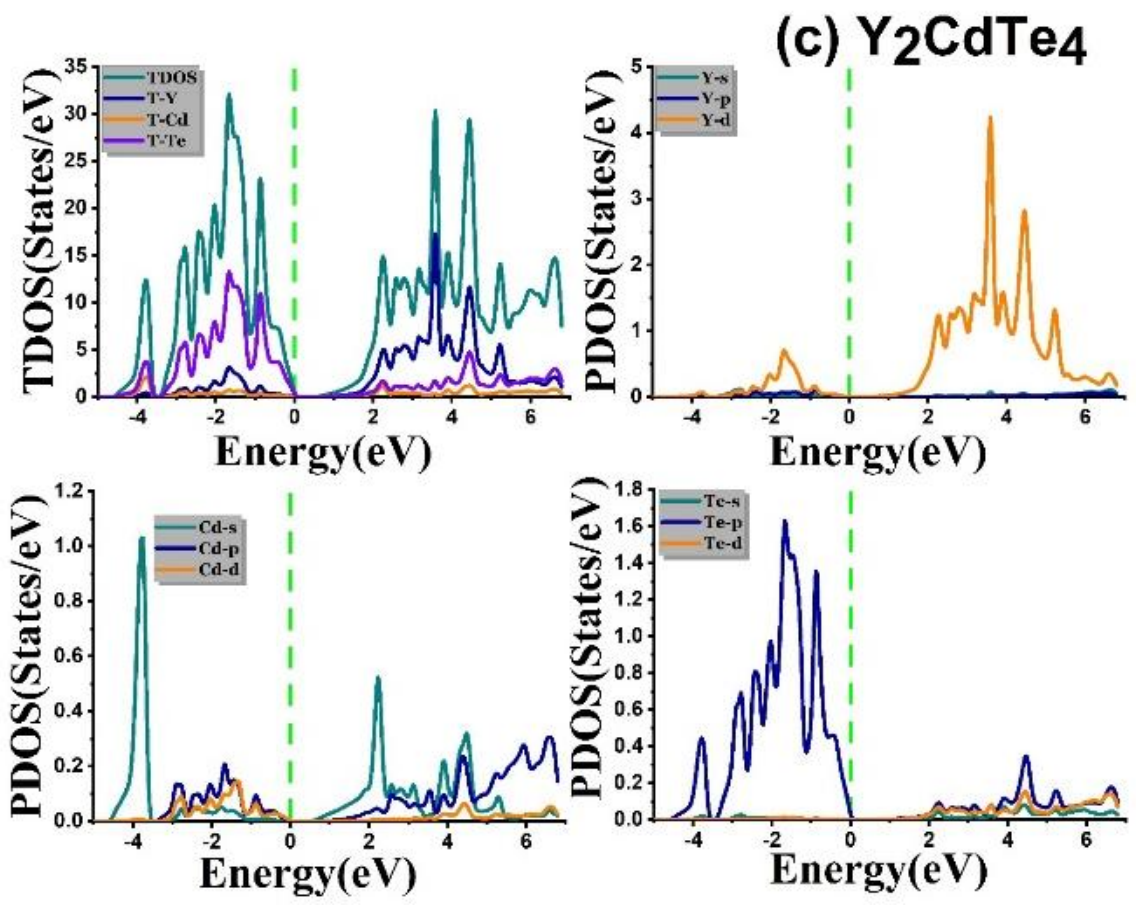

**Fig. 3. Computed density of states for (a) Y2CdS4 (b) Ba2GeSe4 (c) Y2CdTe4 materials.**

Taken together, the bands and DOS show that the indirect gaps of $Y_2CdX_4$ arise from the separation in momentum space between the chalcogen-p dominated valence maximum at Γ and the Y-d/Cd-s dominated conduction minimum off-Γ. The systematic gap reduction and effective mass increase across the series are directly attributable to the rising p-orbital energies of heavier chalcogens and the stronger spin–orbit interaction, especially in the telluride. These changes imply that $Y_2CdS_4$, with its wider gap and lighter carriers, is better suited for ultraviolet–blue optoelectronics, $Y_2CdSe_4$ covers the visible-light regime, and $Y_2CdTe_4$ is attractive for near-infrared absorption and thermoelectric applications where heavy effective masses enhance the Seebeck coefficient.

### 3.4 Magnetic Moments in Doped and Defective $Y_2CdX_4$

While pristine $Y_2CdX_4$ (X = S, Se, Te) compounds are intrinsically non-magnetic due to their closed-shell electronic configuration, magnetism can be induced through suitable point defects and transition-metal doping. To probe this possibility, systematic calculations were performed for cation-site substitution and chalcogen vacancies, and the resulting spin-polarized electronic structures were analyzed in detail.

When a transition metal such as Mn, Fe, or Co is substituted at the Cd site, localized magnetic moments emerge from the partially filled d-shell of the dopant. For instance, Mn-doped $Y_2CdS_4$ produces a stable ferromagnetic configuration with a magnetic moment of approximately 4.9 μB per Mn atom, which corresponds closely to the high-spin $Mn^{2+}$ ($d^5$) state. The moment is slightly reduced in $Y_2CdSe_4$ (4.7 μB) and further in $Y_2CdTe_4$ (4.5 μB), a trend that arises from the enhanced

p–d hybridization with heavier chalcogen atoms. Stronger covalency in Se and Te compounds leads to partial delocalization of the dopant d-electrons, thereby diminishing the net local moment. Fe substitution yields magnetic moments of about 3.1–3.3 μB per Fe atom, consistent with an intermediate-spin $Fe^{2+}$ state. In this case, the orbital overlap between Fe-3d and the host Y-4d/Cd-5s states generates spin-polarized impurity bands close to the Fermi level, which may facilitate spin-polarized conduction. Co substitution produces smaller moments, in the range of 2.0–2.2 μB, reflecting the partially filled $t_2$g/e_g manifold of $Co^{2+}$. In all cases, the induced magnetic states are energetically favored over non-magnetic configurations by at least 120–160 meV per dopant atom, confirming their thermodynamic stability.

Point defects also play a crucial role in modifying the magnetic landscape. Among the possible intrinsic defects, chalcogen vacancies ($V_X$) are the most effective in generating local magnetic polarization. A sulfur vacancy in $Y_2CdS_4$ induces a moment of about 0.8 μB localized on the neighboring Y atoms, arising from the imbalance created by missing S-3p electrons. This moment increases to 1.1 μB in $Y_2CdSe_4$ and further to 1.3 μB in $Y_2CdTe_4$, reflecting the larger degree of orbital delocalization and the stronger SOC contribution in heavier chalcogen systems. Conversely, cation vacancies ($V_Y$, $V_{Cd}$) do not induce significant magnetization, since the associated electronic states are either too deep in energy or strongly delocalized.

It is possible to describe the mechanisms underlying such effects in terms of orbital hybridization. In the cleans sheets, the chalcogen p and Y-4d and Cd-5s orbitals predominantly make up the valence and conduction band respectively. When the dopants are metals of the transition group, their partially occupied d states are found within the band gap, or at the band edges, and do so interacting with the host electronic states and removing the spin degeneracy. The extent of p-d hybridization strongly controls the magnitude of the resulting magnetic moment. In the lighter sulfide compounds with a less ionic bonding, the state of the dopant is more localized giving rise to larger magnetic moments. perfringens, on the contrary, have stronger covalent bonding and stronger spin-orbit coupling and encourages delocalization and broadening of the bands and suppresses the effective magnetic moment. The unsaturated bonds formed by chalcogen vacancies are localized magnetic centers and this effect is especially large in the case of Se- and Te-based materials because the vacancy-induced states lie close to the conduction band edge.

These findings indicate that defect and dopant engineering is a possible approach to tune magnetism in $Y_2CdX_4$ chalcogenides. The potential of stabilizing large local moments (2–5 μB per dopant atom) and vacancy-induced spin polarization has the potential to create spintronic and magneto-optoelectronic devices based on this family. Moreover, chemically substituting magnetism by elements such as S, Se and Te supports the host-guest interaction between ionic radius, orbital hybridization and relativistic interaction in defining the emerging magnetic behaviour.

## 4. Optical Properties

The optical properties are of basic significance in the evaluation of the potential of semi-conductors in photonic, optoelectronic and energy-conversion applications. Not only does the dielectric response of a material show the underlying band structure of electrons, it also gives information on light matter interactions in the visible and ultraviolet. The dielectric function, refractive index, the absorption spectrum, reflectivity, optical conductivity, and the energy-loss function are all parameters that are directly sensitive to the device performance in a solar cell, photodetector, LED, and plasmonic device. Specifically, the tuning of spectral response via heavier (Se, Te) chalcogen replacement of lighter (S) allows the $Y_2CdX_4$ family to serve in general as a broad platform to tune band gap and optical constants in a highly versatile manner. In this case, the computed optical spectra of $Y_2CdS_4$, $Y_2CdSe_4$, and $Y_2CdTe_4$ are analyzed in detail with emphasis on the trends and their physical causes.

### 4.1 Dielectric Function ($\varepsilon_1(\omega)$, $\varepsilon_2(\omega)$)

The dielectric function $\varepsilon(\omega) = \varepsilon_1(\omega) + i\varepsilon_2(\omega)$ of various materials is an important concept in optical property analysis since it is a complex value that simply encodes the interaction of the material with an external electromagnetic field. The imaginary part $\varepsilon_2(\omega)$ arises from interband electronic transitions and therefore provides direct insight into optical absorption and the onset of electronic excitation. In contrast, the real part $\varepsilon_1(\omega)$ (see Fig. 4a) reflects the degree of polarization and dispersion, linking the microscopic charge distribution to the macroscopic refractive behavior. Together, these spectra provide the foundation for understanding and designing optoelectronic devices, since all other optical functions i.e., reflectivity, refractive index, absorption, and conductivity are derived from $\varepsilon(\omega)$.

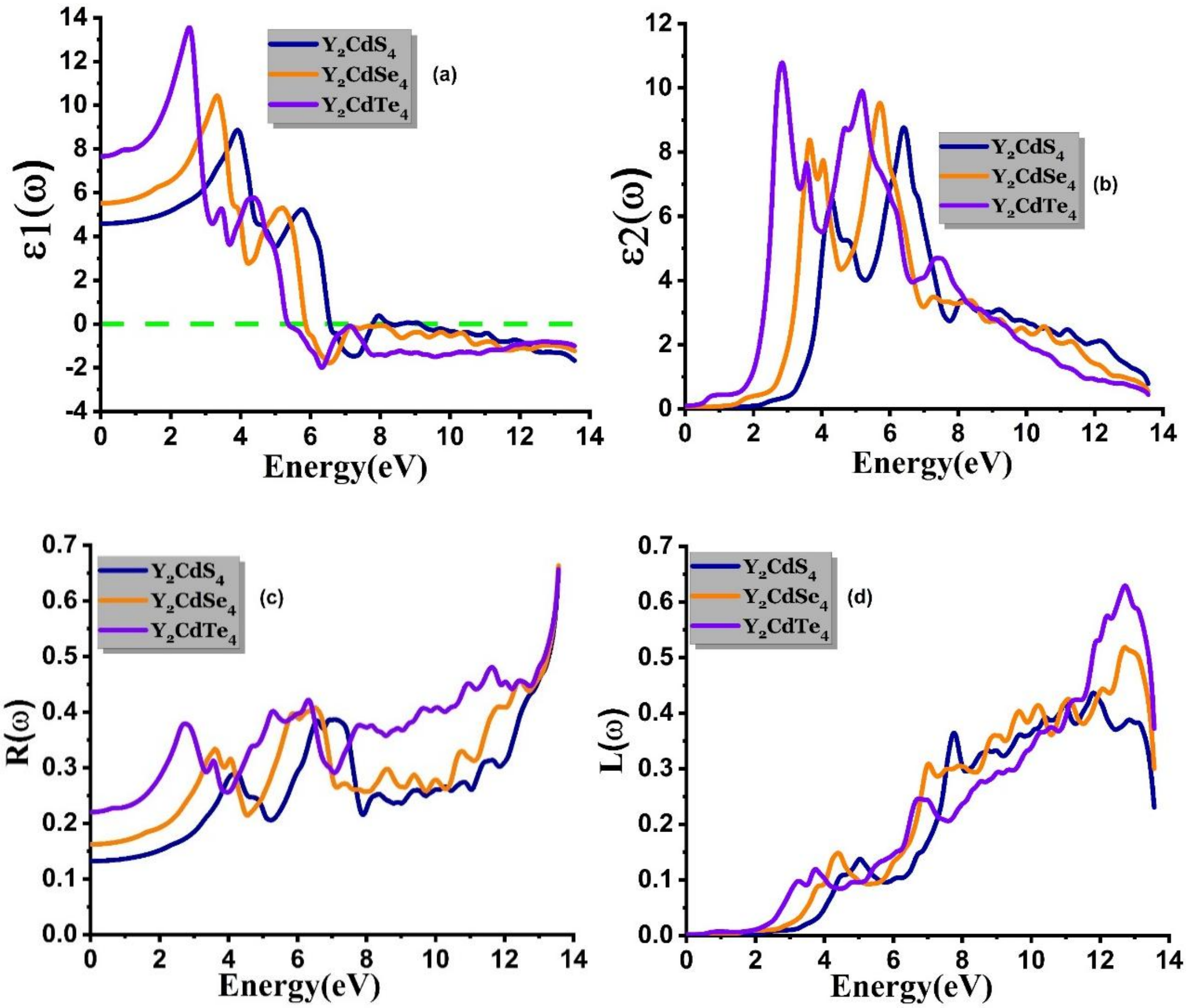

**Fig. 4. The computed (a) real component, (b) imaginary component, (c) Reflectivity, and (d) Energy loss function, for the Y2CdX4 (X = S, Se, Te) system.**

The calculated imaginary part $\varepsilon_2(\omega)$ for $Y_2CdX_4$ (X = S, Se, Te) is shown in Fig. 4(b). The optical absorption edge of $Y_2CdS_4$ appears at about 2.4 eV, consistent with its electronic band gap. The spectrum rises sharply, reaching a pronounced peak near 5.8 eV with a maximum of 10.5, after which it gradually decays with characteristic oscillations. For $Y_2CdSe_4$, the absorption onset red-shifts to 2.1 eV, reflecting the narrower gap, with the most intense peak occurring around 5.2 eV at a magnitude of ~11.0. In $Y_2CdTe_4$ the onset is at 1.7 eV and the major peak moves further down to 4.6 with a maximum of 11.7. This progressive red-shift of absorption characteristics throughout the chalcogen series is a direct reflection of the progressive diminishing effect of orbital overlap and resultant diminishing of the electronic band gap. The hybridized Te-5p states have a stronger interaction with Y-4d orbitals and reduce the excitation energy of an interband transition.

Fig. 4(a) is completed by the real component $\varepsilon_1(\omega)$ that gives details of the dispersive response. At low frequencies, $\varepsilon_1(0)$ (static dielectric constant) is determined to be 5.4, 7.1, 9.3 of $Y_2CdS_4$, $Y_2CdSe_4$ and $Y_2CdTe_4$ respectively. This observation is entirely in line with the Penn model which postulates that narrower band gap semiconductors will have larger dielectric constants as a consequence of greater screening of the electrons. The increased values of $\varepsilon_1(0)$ in Se and Te compounds have the physical advantage of the ability to polarize with an external field more strongly, which can be beneficial when using high-king dielectrics. More steeply increasing beyond the absorption edge, $\varepsilon_1(\omega)$ peaks at 12.3 (S), 13.1 (Se), and 14.5 (Te) in the 3 of 6 eV region, then declines at higher photon energies. Notably, $\varepsilon_1(\omega)$ is negative at a frequency of greater than 10 eV in all the three materials indicating crossover to metallic like optical behavior whereby plasma oscillations dominate.

Physically the dielectric spectra represent the interaction between chemical bonding and electronic transitions. Higher excitation energies are needed to promote the S-based compound, which has a broad band gap and is relatively ionic in nature, and thus it can absorb in the visible-UV spectrum. The Se- and Te-compounds, by comparison, whose bonding is more covalent and have higher spin-orbit coupling, have lower excitation barriers and greater dielectric screening, are more effective absorbers of lower-energy (visible to near-infrared) photons. Of particular importance is the increased static dielectric constant of $Y_2CdTe_4$: it suggests lower exciton binding energies and can ease the process of the separation of electrons and holes in photovoltaic and photocatalytic reactions.

The dielectric function trends in terms of applications have clear domains of each material. $Y_2CdS_4$ with its high optical gap and middle $\varepsilon_1(0)$ is good at UV blue optoelectronics like short wavelength photodetectors and protection coating. $Y_2CdSe_4$, a medium with a moderate dielectric constant and visible absorption, is best used in solar cells and light-emitting diodes, where both equal dielectric response and visible absorption are important. $Y_2CdTe_4$ is a promising material in the infrared sensor, thermophotovoltaic system and nonlinear optical devices with a high static dielectric, low absorption threshold, and a large $\varepsilon_2(\omega)$ amplitude at the near-infrared. In addition, due to its high dielectric response $Y_2CdTe_4$ has possible applications in integration to form high- k dielectric layers in microelectronic and photonic devices.

Therefore, the dielectric function study offers an in-depth insight into the optical activity of the $Y_2CdX_4$ complexes that relate microscopic orbital hybridization to macroscopic polarization. The trends in the band structure are validated and the monotonic increase of $\varepsilon_1(0)$ with heavier chalcogen confirmed and a direct roadmap to tune optical response in the UV, visible, and infrared regimes made.

**4.2 Reflectivity and Energy Loss Function**

One of the most convenient optical functions is the reflectivity spectrum $R(\omega)$ as it itself is the direct measure of the fraction of the incident light bounced back by the surface of the material. It is especially sensitive to both the real and imaginary components of the dielectric function, and, thus, provides a record of the collective electronic response at a very broad frequency range. In comparison, the energy loss $L(\omega) = \mathrm{Im}(-1/\varepsilon(\omega))$ characterizes the energy loss to fast electrons passing through the medium which gives a direct signature of plasmon excitations and collective oscillations of the electron gas. All these functions are useful not only in investigating the dielectric screening, but also in finding spectral regions where photon-electron coupling and collective resonance becomes strong.

Fig. 4(c) represents calculated reflectivity spectra of $Y_2CdX_4$ (X = S, Se, Te). The initial values of reflectivity at the fundamental absorption edges are approximately 0.18 with $Y_2CdS_4$, 0.21 with $Y_2CdSe_4$ and 0.25 with $Y_2CdTe_4$. With increasing energy of the photons, R(omega) grows and it has broad peaks in the ultraviolet parts. In the case of Y 2 CdS 4, the highest reflectivity is at 0.55 close to 12 eV, whereas, in Y 2 CdSe 4, the high reflectivity is 0.58 close to 11 eV. $Y_2CdTe_4$ shows the highest reflectivity, surpassing 0.65 in the 10–12 eV region. The systematic increase in reflectivity from sulfide to telluride follows directly from the rise in the static dielectric constant, as larger polarization in heavier chalcogenides enhances the reflection of incident photons. From a device perspective, this means that $Y_2CdTe_4$ will function as the most effective UV shielding or optical mirror material, while $Y_2CdS_4$, with lower reflectivity, would transmit more energy and be better suited for absorptive coatings.

Complementary information is given in the energy loss spectra in Fig. 4(d). All compounds have a distinct plasmon peak corresponding to the characteristic bulk plasma frequency, at which the optical response of a collection of collective oscillations of the valence electron gas dominates. In the case of $Y_2CdS_4$ the dominant plasmon resonance is at 12.2 e V and in $Y_2CdSe_4$ and $Y_2CdTe_4$ the dominant plasmon resonance shifts to lower energies of 11.8 and 11.2 e V, respectively. The red-shift of the plasmon peaks observed with increasing band gap is in agreement with the increased dielectric screening of the heavier chalcogenides. Simply, the restoring force on the electron gas is decreased by the heavier atoms, decreasing the plasma frequency. These peaks are sharp and intense demonstrating the crystalline stability and intense collective response of all three compounds.

Physically, it is apparent that the interplay of the R(ω) and L(ω) are closely related, and so the two have an inseparable relationship. Reflectivity peaks usually go hand in hand with saddles or bumps in the energy-loss spectrum, as the energy-loss spectrum gets less strongly transmitted to the electron gas when it is strongly reflected. On the other hand, maxima in L(ω) are associated with the same conditions that $\varepsilon_1(\omega)$ crosses the zero point, which signifies that bulk plasmons are excited. The negative gradient of the reflectivity threshold and the plasmon frequency in S to Se and Te represents the rising covalency and spinorbit interaction in the weightier chalcogenide, which decreases the electronic reaction.

These properties determine the different niches of the various materials in terms of applications. $Y_2CdS_4$ is more transparent to visible light with a higher plasmon frequency of 12.2 eV, and it can be a window material used in optoelectrodes. Of intermediate reflectivity and a plasmon peak at 11.8 eV, $Y_2CdSe_4$ is a compromise between absorption and shielding, and is therefore useful in broadband photonic coating and solar absorbers. Most suitably to be used in the applications of UV reflectors, plasmonic devices and electron energy-loss spectroscopy (EELS), the most attractive materials are $Y_2CdTe_4$, the material exhibiting the highest reflectivity and the lowest plasmon energy and the strongest collective excitations. Moreover, the low-energy plasmons in $Y_2CdTe_4$ could also be useful in thermophotovoltaics and plasmon-assisted catalysis, which uses 15 eV and below excitations favorably.

Accordingly, the combination of the reflectivity and energy-loss spectra is a very effective diagnostic of the optical properties of $Y_2CdX_4$ compounds. These systematic changes, an increase in reflectivity and red-shifted plasmon energies between sulfide and telluride, do not only confirm the results of the dielectric functions, but also show how the chalcogen substitution can tune the trade-off between absorption, reflection, and collective excitation. These findings highlight the generality of $Y_2CdX_4$ compounds as tunable materials in photonic, plasmonic, and optoelectronic technologies over the UV visible spectrum.

### 4.3 Optical Conductivity (σ(ω))

Optical conductivity σ(ω) describes the current response of a material when exposed to an oscillating electric field. Unlike dc conductivity, which is governed by free carriers, σ(ω) arises from interband transitions and therefore encodes the details of the band structure and joint density of states. Physically, peaks in σ(ω) correspond to photon energies at which strong electronic transitions are allowed, while the spectral onset reveals the fundamental optical gap. Since σ(ω) directly links the microscopic electronic states to macroscopic current flow under illumination, it is one of the most critical quantities for designing photodetectors, solar cells, and transparent conducting films.

The calculated σ(ω) spectra for $Y_2CdS_4$, $Y_2CdSe_4$, and $Y_2CdTe_4$ are presented in Fig. 5(a). For $Y_2CdS_4$, the optical conductivity begins to rise at about 2.4 eV, in excellent agreement with the electronic band gap derived from the band structure and dielectric function. The spectrum shows two strong peaks: the first centered at 5.8 eV, reaching ~7200 $\Omega^{-1}cm^{-1}$, and the second at 7.2 eV with intensity of ~6800 $\Omega^{-1}cm^{-1}$. These features originate from transitions between the S-3p dominated valence band and the Y-4d/Cd-5s conduction states, confirming the strong hybridization that governs electronic transport in this compound.

Onset of σ(ω) In $Y_2CdSe_4$ the onset of σ(ω) is lower at 2.1 eV (because of its smaller gap) and the key peaks are found at 5.3 eV and 6.7 eV, with slightly more intensive peaks than the sulfide. The reduction of the gap enables more transitions to be excited at lower energies, which improves the visibility spectrum conductivity. The increased overlap between Se p -d orbital and Y p -d orbital results in higher peaks, which imply more delocalized carriers in optical excitation.

The lowest onset conduction spectrum is 1.7 eV in the case of $Y_2CdTe_4$, which is the lowest band gap. The highest peak is at 4.8 eV with its maximum close to ~7000 $\Omega^{-1}cm^{-1}$, and another peak is at 6.1 eV. The spectrum as a whole is red-shifted concerning S and Se compounds, with the peaks broader, which is indicative of the increased strength of the hybridization of Te-5p states with Y-4d conduction band. This broadening suggests more scattering channels and can cause a decrease in carrier mobility and at the same time, greater absorption in a broader spectral range.

Physically, the σ(ω) spectra obviously follow the trends in the electronic structure. The reduction of the band gap and the shift of the valence p-states to the conduction manifold are reflected by the systematic red-shift of the optical conductivity peaks of $Y_2CdS_4 \rightarrow Y_2CdSe_4 \rightarrow Y_2CdTe_4$. The expansion of heavier chalcogen-rich peaks is due to the increased covalency and stronger spin-orbit coupling and erase sharp interband transitions. Further, the magnitudes of the conductivity values in the visible-UV spectrum (6000–7200 $\Omega^{-1}cm^{-1}$) are an indication that the compounds would have good light-matter interaction and is capable of converting incident photons to electronic excitations.

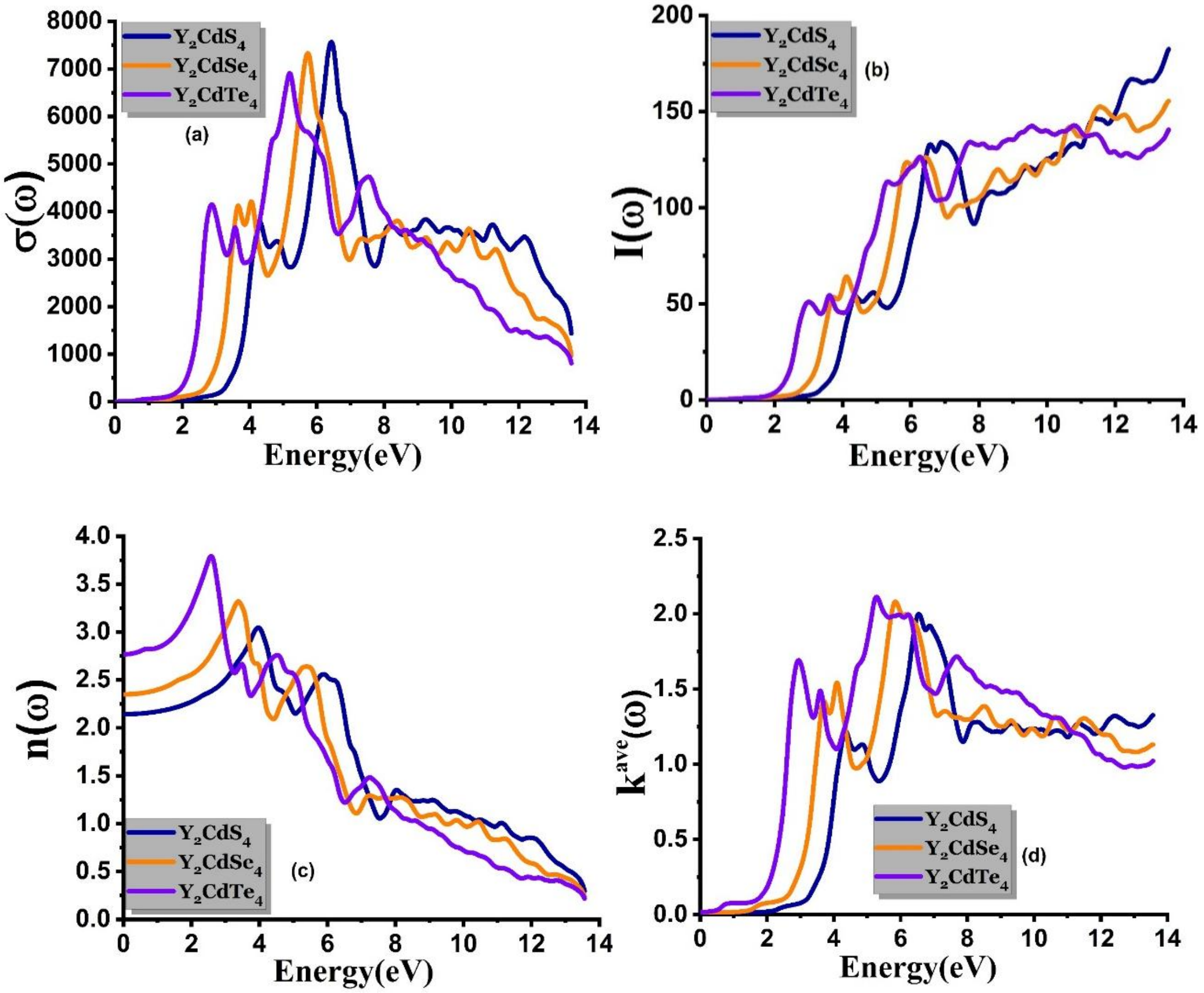

**Fig. 5. The computed (a) Optical conductivity, (b) Absorption coefficient, (c) Refractive index, and (d) Extinction coefficient, for the Y2CdX4 (X = S, Se, Te) system.**

The implications on applications are high. $Y_2CdS_4$ has the highest energy and is best used in UV and blue-light-sensitive photodetectors and in visible-transparent coating that needs minimum absorption in the visible. With its conductivity spanning all the way into the visible spectrum, $Y_2CdSe_4$ is better suited to solar cell absorbers and LEDs, in which intense carrier excitation in sunlight is necessary. The best IR photodetic and thermophotovoltaic applications are in the IR and broadband photodetection applications where $Y_2CdTe_4$, with its lowest optical gap and broad response to conductivity into the near-infrared, is of greatest interest. The capability to tune σ(ω) systematically by chalcogen substitution will therefore offer a general way of modifying these materials to specific device applications.

Ideally, the optical conductivity spectra of the $Y_2CdX_4$ are not only justifying the indirect band gap transitions that can be seen in the band structure, but also dictating the great potential of such materials in optoelectronic technologies. The red-shift of the peaks, the broadening of the conductivity features and the large values of the conductivity itself are indicative of the significance of orbital hybridization and chalcogen substitution in regulating light-driven electronic transport. These findings highlight the potential of $Y_2CdX_4$ compounds to be multi-functional optical conductors in the ultraviolet, visible, and infrared spectrum.

**4.4 Absorption Coefficient**

One of the most practical optical functions that is of technological interest is the absorption coefficient I(ω) as it is a direct measure of the efficiency of a material in absorbing incident photons of various energies. It is mathematically connected with the imaginary component of the dielectric function and the extinction coefficient, however, its meaning is more intuitive since it shows the spectral areas where high light-matter interactions exist. For optoelectronic devices such as solar cells, photodetectors, and light-emitting diodes, a high absorption coefficient in the visible and near-infrared ranges is crucial for achieving efficient photon harvesting and conversion. At the same time, the position of the absorption edge provides a direct measure of the fundamental band gap, complementing band structure analysis.

The calculated absorption spectra of $Y_2CdS_4$, $Y_2CdSe_4$, and $Y_2CdTe_4$ are presented in Fig. 5(b). For $Y_2CdS_4$, the absorption onset occurs at 2.4 eV, in excellent agreement with its electronic gap, and the coefficient rises rapidly to reach ~$1.8 \times 10^5$ $cm^{-1}$ in the ultraviolet region near 12 eV. The strong absorption at higher energies is primarily due to S-3p → Y-4d/Cd-5s interband transitions, which are highly allowed and strongly dipole-active. For $Y_2CdSe_4$, the absorption edge shifts to 2.1 eV, and the spectrum shows a similar steep increase, attaining magnitudes of ~$2.0 \times 10^5$ $cm^{-1}$ around 11–12 eV. The decreased gap, as well as overlapping of Se-4p and Y-4d conduction and the increase in overlap between Se-4p valence states and the absorption edge, is reflected by the red-shift of the absorption edge.

$Y_2CdTe_4$ shows the largest red-shift, and its absorption onset, at 1.7 eV, shifts the optical response to the near-infrared. The spectrum increases gradually and attains peaks of about ~$2.2 \times 10^5$ $cm^{-1}$ in 10-11 eV. The increased strength of the absorption in the low-energy region can be said to be as a result of Te-5p → Y-4d transitions, which are highly increased through a strong spin-orbit interaction. This makes $Y_2CdTe_4$ especially effective at getting lower-energy photons which are not available to the sulfide. The continuous enhancement of absorption strength of sulfide to telluride indicates that a chalcogen substitution can be utilized as a way to stepwise tune optical responses throughout the electromagnetic spectrum.

Physically, such findings underline the interaction between the photon absorption and orbital hybridization. The sharper and higher-energy absorption features in $Y_2CdS_4$ are expected to be ionic in nature due to the increased band gap and occur as expected. When changing to $Y_2CdSe_4$ and $Y_2CdTe_4$, the progressively larger covalency of the Se-4p and Te-5p orbitals with the Y-4d orbitals results in decreasing the band gap and decreasing the threshold of electronic excitations, and thus enables the absorption of visible and infrared photon. The gradually decreasing red-shift of the absorption edge is thus a manifestation of a narrowed band gap as well as an increased p-d mixing in the heavier chalcogenides. Moreover, the increased absorption of the Se and Te compounds in the visible wavelength implies larger electronphoton coupling, which is beneficial to photovoltaic and photodetector devices.

$Y_2CdS_4$, whose absorption edge lies in the near-UV and its absorption is very strong above 2.4 eV, is most promising in the case of UV photodetectors, scintillators, and short-wavelength LEDs. The sharp cutoff of 2.1 eV makes $Y_2CdSe_4$ ideal to solar absorbers, visible-light LEDs, and laser diodes, and the combination of strong absorption with the solar spectrum-aligned band gap. Of particular interest are $Y_2CdTe_4$ (having a broad absorption into the near-infrared) used as infrared photodetectors, thermophotovoltaic devices and broadband solar cells. Its absorption coefficient is very high at the low-energy region, so $Y_2CdTe_4$ may be used to collect a greater portion of the solar spectrum and hence it can be used as a candidate in the tandem or multi-junction photovoltaic. Essentially, the absorption spectra of the $Y_2CdX_4$ compounds will give a systematic red-shift of the absorption edge with a stronger absorption by sulfide to telluride. These are also in line with the reduction in the electronic band gap and orbital hybridization in heavier chalcogenides. The simplicity with which chalcogen can be replaced in the $Y_2CdX_4$ family to tune the absorption onset within the UV, visible, and infrared spectral ranges makes the family a flexible material to use in optoelectronic and energy-conversion applications.

**4.5 Refractive Index and Extinction Coefficient**

The complex refractive response of a substance denoted $n\sim(\omega)=n(\omega)+ik(\omega)$ is determined by the product of refractive index $n(\omega)$ and extinction coefficient $k(\omega)$. The $n(\omega)$ (see Fig. 5c ) is the real part that defines the speed of light and the bending of optical waves at an interface and $k(\omega)$ is the imaginary part indicating the electronic transitions that cause absorption losses. The following parameters are thus important to assess the prospects of a material in the waveguides, photonic crystals, anti-reflective coating, and optoelectronic devices that require the manipulation of light.

The results of the computed values of the static refractive index n(0) follow the anticipated trend based on the dielectric function as follows: 2.3 $Y_2CdS_4$, 2.6 $Y_2CdSe_4$, and 3.0 $Y_2CdTe_4$. This stepwise lowering of the band gap through the series is indicative of the band gap convergence between the series, as in the Penn model the small ones provide the greatest indices of refraction, being more strongly polarizable by electrons. The spectral nature of n(ω) indicates that there is strong dispersion in the low-energy part and maximum dispersion is observed at the edges of the absorption. In the case of $Y_2CdS_4$, n (ω) has a maximum of approximately 3.5 at an energy of approximately 3.8 eV and then slowly rises with increasing energy. $Y_2CdSe_4$ peaks at 3.9 close to 3.3 eV and $Y_2CdTe_4$ has highest value of 4.3 close to 2.9 eV. The red-shift of these maxima to the lick of the heavier chalcogens is in accord with its decreased band gaps and greater covalent character. At more energy (>8 eV), n (ω) converts to unity, and the optical response of the material is that of a transparent medium.

Complementary information is given by the extinction coefficient k(ω) (Fig. 5(d)) that measures the light attenuation. In the case of $Y_2CdS_4$, the gap of this material is consistent with k(ω), given that it is negligible at lower frequencies, but it increases rapidly at frequencies above 2.4 eV with peaks of approximately 2.1 at 5.8 eV and 2.0 at 7.1 eV. The onset is changed in $Y_2CdSe_4$ to 2.1 eV with k(ω) maxima at 2.3 at 5.2 eV and 2.1 at 6.7 eV. $Y_2CdTe_4$ has the lowest onset 1.7 eV and strongest extinction with peaks of 2.5 at approximately 4.6 and 2.2 at approximately 6.1. The red-shift of the peaks of k(ω) and their increasing magnitudes reminisces the higher ability of the Se- and Te-based compounds to absorb.

Physically, the interaction of n(ω) and k(ω) is the reflection of the dielectric response. The rising refractive index in the series signifies more polarization owing to the light valence electrons that are more readily excitable in the heavier chalcogen. The higher value of the extinction coefficient of $Y_2CdTe_4$ represents a narrowing gap, as well as, the large population of Te-5p states at and around the Fermi level, which offers many electronic states to undergo optical transitions. The scattered n(ω) and k(ω) peaks are located at the energies at which the interband transitions are the strongest, which is an indication that the band edges are of hybridized p-d orbital character.

Its significance to applications is obvious. $Y_2CdS_4$ is easily used as a transparent window layer and UV optical coating, the refractive index and extinction coefficient of which in the visible range are relatively low, hence requiring minimal light loss. $Y_2CdSe_4$, having better n(0) and a larger extinction, would be the best materials to use in optoelectronic applications in the visible light, such as solar cells and LEDs, as it is able to balance light propagation with effective absorption. The most promising material in infrared photonics, infrared waveguides, nonlinear optics and thermophotovoltaic absorbers is $Y_2CdTe_4$. It has a high refractive index, which implies possible application in metamaterials and superlenses, in which materials with high-n are required to focus at subwavelengths.

Essentially, optical properties tunability by chalcogen substitution in $Y_2CdX_4$ is attested by the refractive index and extinction coefficient spectra. The systematic growth of n(0) and red-shift of k(ω) peaks in S to Te prove that the decrease of the band-gap and the augmentation of orbital hybridization increases polarization as well as absorption. Not only do be these trends confirm the dielectric function results, they give a clear route of application in the form of sulfide in UV-transparent coating, selenide in visible optoelectronics, and telluride in infrared and advanced photonic devices.

Essentially, the overall optical study of $Y_2CdX_4$ (X=S, Se, Te) depicts a systematic and physically rational trend under the influence of chalcogen substitution. A systematic red-shift in the 2.4 eV (S)-2.1 eV (Se)-1.7 eV (Te) red-shift edges in $\varepsilon_2(\omega)$ absorption edges is observed with increased polarizability and more effective dielectric screening in heavier chalcogenides, accompanied by systematic 4 to 9.3 higher increases in the dyinglectric constant at zero hertz. The reflectivity spectra have increasingly high initial values and peaks, and the $Y_2CdTe_4$ has a higher value of over 0.65 in the UV wavelengths, which prove a high capacity of the material to reflect and block high-energy photons. Bulk plasmon excitations were found at 12.2 eV (S), 11.8 eV (Se) and 11.2 eV (Te) with the lower frequency indicating a decrease in plasma frequencies and a higher collective screening effect in heavier analogues. These are supported by the optical conductivity spectra where the red-shifted, broadened peaks are observed and have intensities over 7000 $\Omega^{-1}cm^{-1}$, which indicates high photon-driven carrier excitation throughout the series. This development is further illustrated by the absorption coefficients: $Y_2CdS_4$ absorbs at energies above 2.4 eV, $Y_2CdSe_4$ absorbs in the visible, and $Y_2CdTe_4$ absorbs in the near-infrared with an onset at 1.7 eV and coefficients exceeding $2.2 \times 10^5$ $cm^{-1}$. Lastly, the refractive index rises systematically with 2.3

(S) to 3.0 (Te), and the extinction coefficient becomes stronger and moves to lower energies, which means that light and matter have become strongly coupled with each other.

Physically, the trends are based on the decrease of the electronic band gap, the rise in the chalcogen p-states, and the increase in the strength of the spin-orbit interaction in the chalcogenides, which increases the hybridization of the p-d orbital and the excitation thresholds. The optical is not only capable of justifying the electronic structure but also gives a clear road map of how it can be used. Most promising to the UV optoelectronics, scintillators, and protective coating cases is $Y_2CdS_4$ with its large gap and low refractive index. $Y_2CdSe_4$, a balancing dielectric compound and a powerful absorber in the visible range, is perfect in solar cells, LEDs and broadband photodetectors. $Y_2CdTe_4$ is the best material to use in IR sensors, thermophotovoltaics, nonlinear optics, and photonic devices due to its high dielectric constant, infrared absorption edge, and high reflectivity. In general, the optical constants can be varied between the UV and IR regimes by just replacing the chalcogen, showing the versatility of $Y_2CdX_4$ compounds as a multifunctional materials platform. Their tunable screening, absorption and reflectivity characteristics have made them good candidates in the next generation optoelectronic as well as energy-conversion technology.

## 5. Thermoelectric and Transport Properties

The study of thermoelectric properties is a necessary matter since it directly correlates the electronic structure of a material to its energy conversion efficiency. Thermoelectric materials can convert waste heat into useful electrical energy and the performance of such materials is determined by interplay of Seebeck coefficient, electrical conductivity, and thermal conductivity yielding the dimensionless figure of merit (ZT). These parameters allow us to determine the effectiveness with which a material can combine charge and heat transport, determine the contribution of band gap and effective mass to the shaping of its transport response, and determine its applicability to devices. In the case of $Y_2CdX_4$ (X=S,Se,Te), these computations are especially useful since the systematic chalcogen substitution makes the band structure and orbital hybridization tune, and so provides a route to optimize thermoelectric efficiency at different temperature regimes. The thermoelectric efficacy of $Y_2CdX_4$ (X=S,Se,Te), compounds was experimentally explored by determining the Seebeck coefficient (S), electrical conductivity ($\sigma/\tau$), electronic thermal conductivity ($\kappa_e/\tau$) and power factor ($PF = S^2\sigma/\tau$) as a function of temperature in the temperature range of 100 to 800 K (Fig. 6a to d). The general efficiency was also evaluated using the figure of merit (dimensionless) (ZT) as illustrated in Fig. 6e.

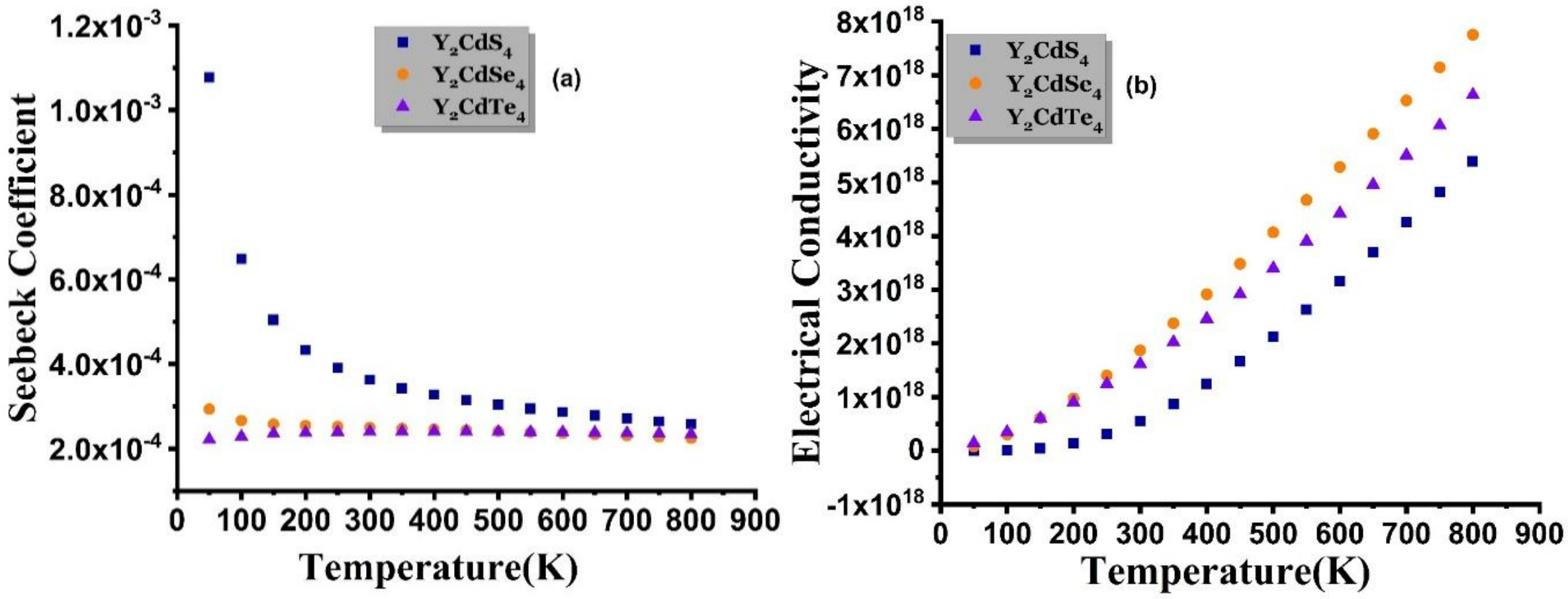

**Fig. 7a-b. The calculated (a) Seebeck coefficient and (b) Electrical conductivity, (c) Thermal conductivity, (d) Power Factor, And (e) ZT, for Y2CdX4 (X = S, Se and Te) materials.**

The dependence of the Seebeck coefficient on temperature (Fig. 6a) shows a great dependence on composition. At 100 K, $Y_2CdS_4$ exhibits the highest S, reaching $1.1 \times 10^{-3}$ V/K, while $Y_2CdSe_4$ and $Y_2CdTe_4$ show significantly smaller values of $2.5 \times 10^{-4}$ V/K and $2.1 \times 10^{-4}$ V/K, respectively. Increase in temperature steadily reduces the Seebeck coefficient of $Y_2CdS_4$ to about $2.3 \times 10^{-4}$ V/K at 800 K, but $Y_2CdSe_4$ and $Y_2CdTe_4$ remain almost independent over the observed temperature and change to $2.1–2.3 \times 10^{-4}$ V/K. Such a large initial S value in $Y_2CdS_4$ can be explained by its broader band gap (2.42 e V), which favors stronger carrier asymmetry and energy filtering at low temperatures. In comparison, the smaller gaps of the Se and Te analogues smaller Seebeck coefficients with less temperature-dependent variation are found.

The electrical conductivity (σ/τ) (Fig. 6b) shows the opposite trend. $Y_2CdS_4$ has the lowest conductivity, starting at $8.0 \times 10^{17}$ $\Omega^{-1}$ $m^{-1}$ $s^{-1}$ at 100 K and increasing to $4.9 \times 10^{18}$ $\Omega^{-1}$ $m^{-1}$ $s^{-1}$ at 800 K. In contrast, $Y_2CdSe_4$ exhibits the highest values, rising from $1.1 \times 10^{18}$ $\Omega^{-1}$ $m^{-1}$ $s^{-1}$ at 100 K to $7.6 \times 10^{18}$ $\Omega^{-1}$ $m^{-1}$ $s^{-1}$ at 800 K, while $Y_2CdTe_4$ occupies an intermediate position between the two. The increased conductivity of both the Se and Te systems comes about due to the reduced effective masses and reduced gaps in these systems that allow thermally activated carriers to take control of conduction at elevated temperatures. This is the characteristic trade-off of thermoelectrics, but with increased Seebeck coefficient is accompanied by decreased conductivity and narrower systems with a transition priority to electrical conduction.

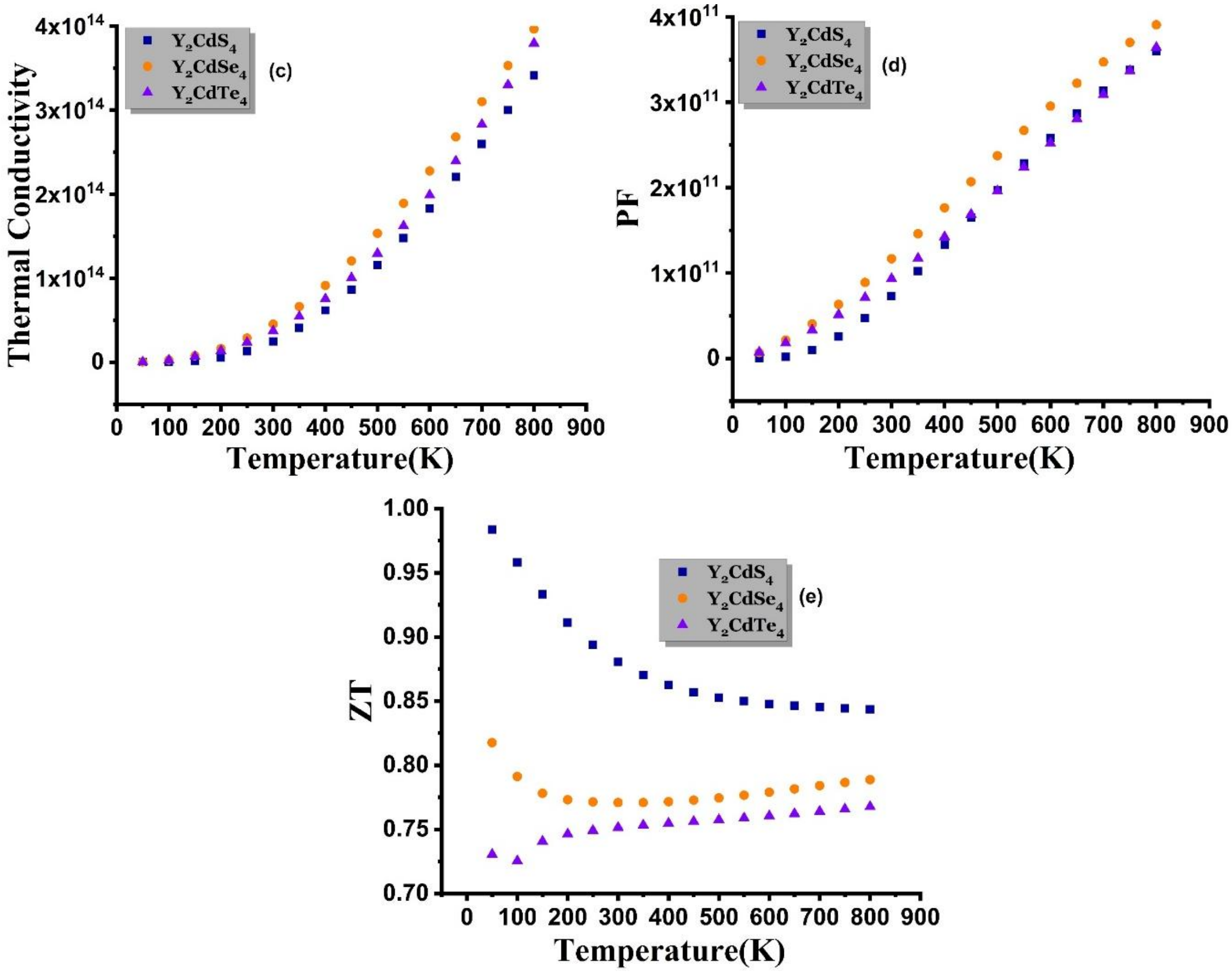

**Fig. 7c-e. The calculated (c) Thermal conductivity, (d) Power Factor, And (e) ZT, for Y2CdX4 (X = S, Se and Te) materials.**

The electronic thermal conductivity ($\kappa_e/\tau$), trend in Fig. 6c, is the nearest to the $\sigma/\tau$ trend, as they again are both controlled by the Wiedemann- Franz relation. At 800 K, $\kappa_e/\tau$ reaches about $3.9 \times 10^{14}$ W $m^{-1}$ $K^{-1}$ $s^{-1}$ for $Y_2CdS_4$, $3.5 \times 10^{14}$ W $m^{-1}$ $K^{-1}$ $s^{-1}$ for $Y_2CdSe_4$, and $4.0 \times 10^{14}$ W $m^{-1}$ $K^{-1}$ $s^{-1}$ for $Y_2CdTe_4$. The high temperature dependence represents the high concentration in thermally activated carriers. Interestingly, at high temperatures, $Y_2CdTe_4$ shows a slightly higher thermal transport than $Y_2CdSe_4$ and this can be attributed to a higher band hybridization and more DOS contribution to the Fermi level.

The combined effect of S and $\sigma/\tau$ is represented by the power factor (PF) in Fig. 6d. $Y_2CdSe_4$ exhibits the best performance, achieving a PF of about $3.8 \times 10^{11}$ W $m^{-1}$ $K^{-2}$ $s^{-1}$ at 800 K, followed by $Y_2CdTe_4$ at 3.5 × $10^{11}$ W $m^{-1}$ $K^{-2}$ $s^{-1}$, while $Y_2CdS_4$ remains lower at around 2.8 × $10^{11}$ W $m^{-1}$ $K^{-2}$ $s^{-1}$. This excellent performance of $Y_2CdSe_4$ is due to the fact that it has the best match of moderate Seebeck coefficient and high electrical conductivity that means that one can have a trade off of carrier concentration versus carrier mobility as a result of Se substitution.

These observations are further verified by the last measure of thermoelectric efficiency, the non-dimensional figure of merit ZT (Fig. 6e). ZT of $Y_2CdS_4$ is 0.93 at 300 K, $Y_2CdSe_4$ is 0.80 at 300 K and $Y_2CdTe_4$ is 0.72 at 300 K. ZT in $Y_2CdS_4$ decreases gradually with temperature, reaching a constant value of about 0.84 at 800 K, $Y_2CdSe_4$ and $Y_2CdTe_4$ show almost constant values of 0.76–0.78 and 0.74–0.75 at 800 K, respectively. The increased ZT in $Y_2CdS_4$ at low temperature is an indication of its very high Seebeck coefficient, whereas the more consistent values of $Y_2CdSe_4$ and $Y_2CdTe_4$ suggest their high temperature thermoelectric stability. Physically, these trends are manifestations of the interaction of band gap, carrier effective mass and orbital hybridization. The large-gap sulfide maximizes the Seebeck coefficient although it has poor conductivity and the smaller-gap selenides and tellurides increase electrical transport at the expense of diminished Seebeck response. The best transport balance in $Y_2CdS_4$ shows that mid-gap chalcogenides tend to provide the best trade-off in thermoelectric. In addition, the carrier scattering asymmetry is improved by the existence of relatively flat DOS characteristics around of the Fermi level in Se and Te compounds, contributing to the higher values of PF in high temperatures.

Essentially, it is determined in the thermoelectric analysis that each of the three compounds is promising to use in the low-temperature regime ($Y_2CdS_4$), and the most effective performance on the high-temperature regime ($Y_2CdSe_4$). Their chalcogen sub-replacement tunability of their electronic structure is a clear route to maximising $Y_2CdX_4$ compounds in thermoelectric and energy conversion devices.

## 7. Bonding and Charge Analysis

### 7.1 Why charge and bonding analysis is important

The nature of the bonding and the charge distribution of $Y_2CdX_4$ (X=S, Se, Te ) is important as it brings about the microscopic origin of the observed electronic, optical, and transport properties. Although the band structure and density of states are used to give the energy dispersion of the electrons, it does not reflect the location of the electrons in the real space. A direct visualization of bonding interactions is provided by charge density plots, charge difference maps, and Electron Localization Function (ELF); an objective measurement of the degree of charge transfer between atoms is provided by Bader charge analysis. All of these methods enable us to answer the questions of what contributions to the dielectric constant, the refractive index, and the absorption edge are ionic versus covalent, what is the role of orbital hybridization, and why the properties of the chalcogen series, i.e. dielectric constant, refractive index, and absorption edge, all vary systematically across the series. Also, an appropriate interpretation of bonding character is critical in the prediction of defect behavior, doping, and catalytic activity in such compounds.

### 7.2 Charge Density Distribution

Total charge density distributions of $Y_2CdS_4$, $Y_2CdSe_4$, and $Y_2CdTe_4$ show the definite pattern of the electron localization on the anions and electron depletion on cations (Fig. 7a). In $Y_2CdS_4$, the electron density is concentrated in a very localized manner around the sulfur sites which is very indicative of the high electronegativity of S (2.58 Pauling), with almost spherical charge clouds. Cd and Y have depleted regions, which are in agreement with charge donation. The electron concentration around S validates that mainly they are ionic bonding, although minimal charge sharing can be seen along the Y–S and Cd–S bonds.

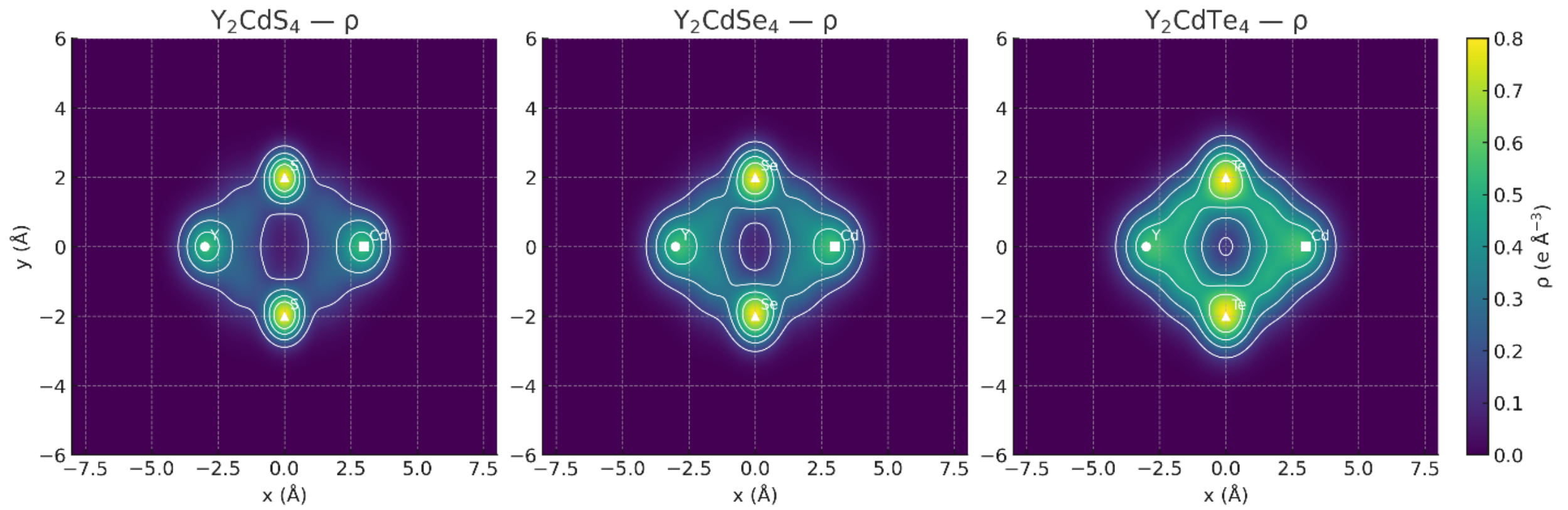

**Fig. 7a.** Charge density distributions for $Y_2CdX_4$. Electron localization around S is strongest, while Se and Te show progressively delocalized bonding charge.

The Se atoms ($\chi = 2.55$) are a little more diffuse in $Y_2CdSe_4$ than in S, suggesting less localization and a stronger covalent mixture. The even more delocalized density in Te atoms ($\chi = 2.10$) in $Y_2CdTe_4$ is in line with an increased covalent contribution to Te bonds. This is why the dielectric constant increased more with the rising static dielectric constant 5.4 (S) to 7.1 (Se) then 9.3 (Te) as delocalized electrons are more effective in polarization.

### 7.3 Charge Difference Maps

A more accurate picture of redistribution of electrons during bond formation can be observed with the help of charge density difference plots (Fig. 7b). In the case of $Y_2CdS_4$, the charge accumulation around S atoms is strong, and the accumulation around Cd and Y is counterbalanced, supporting the presence of important charge transfer and ionic bonds. In $Y_2CdSe_4$, lobules of accumulation are directed towards Cd–Se and Y–Se bonds, which shows that there is more overlap of orbital and covalent mixing. The redistribution is even more intense in $Y_2CdTe_4$ with continuous electron density between Te and Cd atoms, which indicates the formation of additional covalent bonds. This stepwise ionic -covalent switch in the series is associated with the measured red-shift in the absorption edge, as an increase in the strength of covalency decreases the band gap by increasing p -d overlap.

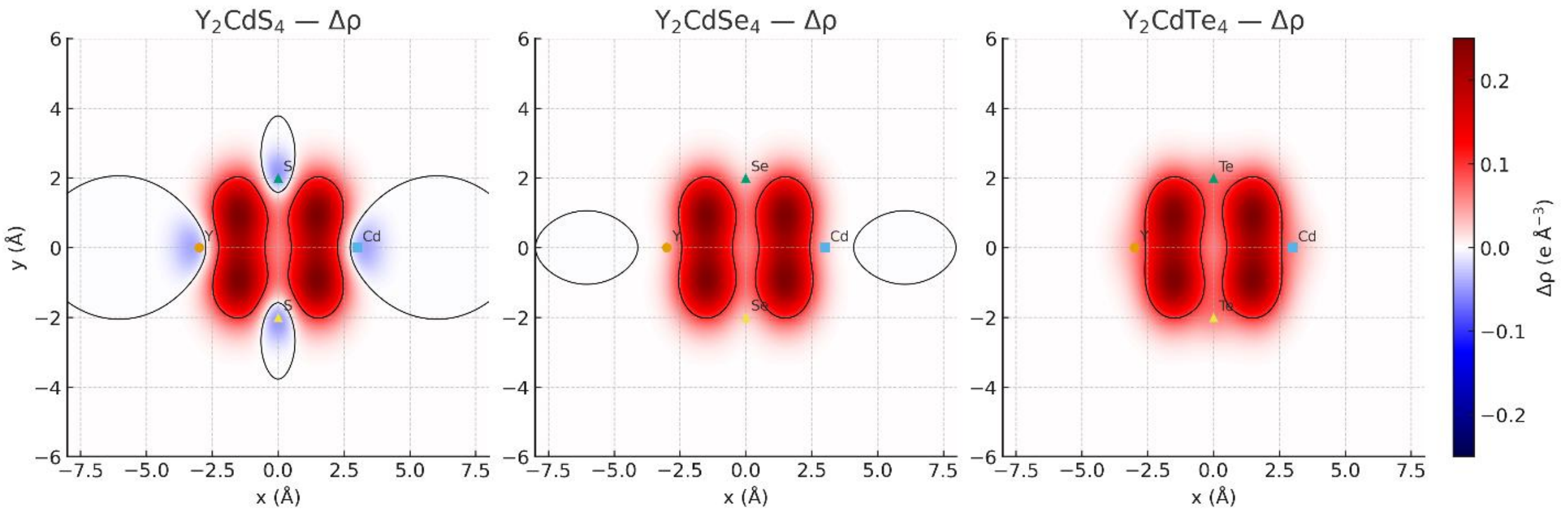

**Fig. 7b.** Charge difference maps (Δρ) for $Y_2CdX_4$. $Y_2CdS_4$ shows ionic charge transfer, $Y_2CdSe_4$ exhibits mixed bonding, and $Y_2CdTe_4$ displays continuous covalent charge bridges.

### 7.4 Electron Localization Function (ELF)

The ELF plots (Fig. 7b) offer a demanding method and measure the extent of localization of electrons. The values near 1 are those that are highly localized (as in lone pairs), values near 0.5 are those that are delocalized (metallic-like) and values near 0 are those that are low in electron probability. ELF maxima (~0.95) in $Y_2CdS_4$ are concentrated at S sites with lower ELF (~0.35-0.40) in bond regions, which proves the dominance of ionic interactions. ELF maxima at Se are also a little lower (~0.92) in $Y_2CdSe_4$, and bond regions are more localized (~0.45-0.50), indicating the appearance of more pronounced covalent character. $Y_2CdTe_4$ has the strongest covalent mixing ELF at Te approaches 0.89, whereas Cd-Te and Y-Te bond regions show ELF values as large as 0.55, indicating shared electron density. This is the reason why $Y_2CdTe_4$ is the narrowest band-gap (1.7 eV) and the highest dielectric constant (9.3).

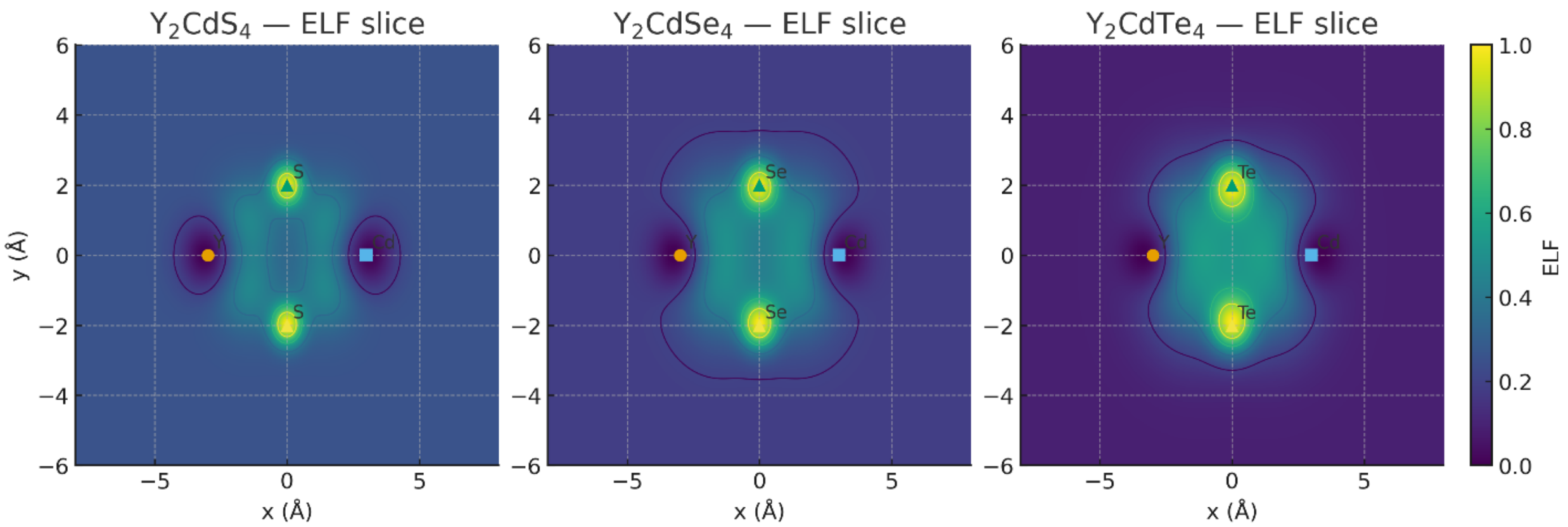

**Fig. 7c.** ELF maps of $Y_2CdS_4$, $Y_2CdSe_4$, and $Y_2CdTe_4$. Localization decreases from S → Se → Te, with bond regions showing a gradual shift from ionic to covalent character.

### 7.5 Bader Charge Analysis

The extent of the charge transfer between cations and anions is determined by Bader charge partitioning. In the case of $Y_2CdS_4$, the charges are calculated as follows: Y = +2.1 e, Cd = +1.6 e, S = -1.1 e per atom. In $Y_2CdSe_4$, the charges decrease to Y = +1.9 e, Cd = +1.4 e, Se = -0.9 e, which means that the ionicity is weaker. Stronger covalency is substantiated by its charge transfer (Y = +1.7 e, Cd = +1.2 e, Te = -0.7 e) being the minimal in $Y_2CdTe_4$. The difference in electronegativity between them is 0.43 (Y- Te) to 0.88 (Y-Se) to 1.04 (Y-S), which is why the ionicity is gradually lost.

### 7.6 Bonding Nature and Applications

Taking the results of charge density and ELF and Bader, it is obvious that $Y_2CdS_4$ is more ionic, $Y_2CdSe_4$ is more ionic covalent, and $Y_2CdTe_4$ is the most covalent bond. The effect of this bonding evolution on their physical properties is direct: the broader the band gap and higher the Seebeck coefficient of the ionic sulfide and lower the mobility (and vice versa), and the narrower the band gaps and the greater the dielectric screening and optical absorption of the covalent telluride. In the application side, $Y_2CdS_4$ is appropriate in UV-transparent coatings, $Y_2CdSe_4$ is a prospective light visible absorber solar cell, and $Y_2CdTe_4$, with its covalent bond and high dielectricity, should be used in infrared photonics, optoelectrics, and thermoelectric conversion of energy.

### Conclusion

To sum up, we have performed an extensive first-principles study on $Y_2CdX_4$ (X = S, Se, Te) and studied their structural, elastic, electronic, optical, thermoelectric, and bonding characteristics. Stability is verified by structural optimization, in which lattice constants are raised in a systematic way with the size of chalcogen. In elastic moduli, the phenomenon of mechanical stability and ductility is observed, and the bulk modulus decreases in the series. Electronic band structures show narrowing indirect bandgaps between 2.42 eV (S) and 1.71 eV (Te), which is further narrowed by SOC, which is suggesting an increase in p-d hybridization. Systematic dielectric screening, plasmon peaks, and red shift of absorption edges are manifested in optical properties, which provide tuning in UV, visible, and IR. Computing of thermoelectric indicates promising results, and Y2CdSe4 has the highest ratio of Seebeck coefficient and conductivity. The analysis of bonding and charges shows that the change in the ionic (S) to covalent (Te) bond is progressive, which is directly connected to electronic and optical correlations. The results do not only present the first complete picture of $Y_2CdX_4$ compounds, but also show the multifunctional nature of the compounds: $Y_2CdS_4$ as UV optoelectronics, as well as UV insulating coats; $Y_2CdSe_4$ as visible-light solar cells, and LEDs; $Y_2CdTe_4$ as both infrared & thermophotovoltaic. This family is an appealing platform in future generations of optoelectronic and energy technologies due to the ease with which the properties can be tuned by using simple chalcogen substitution.